\documentclass[preprint,aps, preprintnumbers, nofootinbib]{revtex4}
\usepackage[dvips]{graphics}

\newcommand{\lsim}[1]{
\setlength{\unitlength}{12pt}
\begin{picture}(1.4,1.)
\put(.7,-0.3){\makebox(0.0,1.)[t]{$<$}}
\put(.7,-0.3){\makebox(0.0,1.)[b]{$\sim$}}
\end{picture}#1}
\newcommand{\gsim}[2]{
\setlength{\unitlength}{12pt}
\begin{picture}(1.4,1.)
\put(.7,-0.3){\makebox(0.0,1.)[t]{$>$}}
\put(.7,-0.3){\makebox(0.0,1.)[b]{$\sim$}}
\end{picture}#2}
\begin{document}
%\preprint{IRB-TH-20/05}
%\preprint{UB-ECM-PF 05/07}
% \draft

\title{Nonsaturated Holographic Dark Energy}

\author{B. Guberina\footnote{guberina@thphys.irb.hr}}
\affiliation{\footnotesize Rudjer Bo\v{s}kovi\'{c} Institute,
         P.O.B. 180, 10002 Zagreb, Croatia}

\author{R. Horvat\footnote{horvat@lei3.irb.hr}}
\affiliation{\footnotesize Rudjer Bo\v{s}kovi\'{c} Institute,
         P.O.B. 180, 10002 Zagreb, Croatia}

\author{H. Nikoli\' c\footnote{hrvoje@thphys.irb.hr}}
\affiliation{\footnotesize Rudjer Bo\v{s}kovi\'{c} Institute,
         P.O.B. 180, 10002 Zagreb, Croatia}

\begin{abstract}
It has been well established by today that the concept of holographic
dark
energy (HDE) does entail a serious candidate for the dark energy of the
universe. Here we deal with
models where the holographic bound
for dark energy is
not saturated
for a large portion of the history of the universe. This is
particularly compelling when the IR cutoff is set by the Hubble scale, since
otherwise a transition from a decelerated to an accelerated era cannot
be obtained for a spatially flat universe. We demonstrate by three generic
but disparate dynamical models, two of them
containing a variable Newton constant, that transition between the two eras
is always obtained for the IR cutoff in the form of the Hubble scale and
the nonsaturated HDE. We also give arguments of why such a
choice for the dark energy is more consistent and  favored over the widely
accepted saturated form.
\end{abstract}
%\pacs{97.60.Gb, 14.60 Pq, 04.80.Cc}
\newpage

\maketitle

After the remarkable discovery of dark energy starting in 1997, a long-standing
and hard-pressing problem in modern
physics called the cosmological constant (CC) problem \cite{1} has become
even aggravated by splitting itself into three distinct (but related ) problems.
Now, besides the `old' CC problem one should also explain why the CC is
small but nonzero (now called the `new' CC problem) as well as 
the near coincidence
of the CC energy density ($\rho_{\Lambda }$) towards the matter energy
density ($\rho_m $)
in the universe at present (the `cosmic coincidence problem') \cite{2}. It
seems today that the most convinced declination of these three problems lies
in the `environmental' variable CC approach based on the string theory
landscape \cite{3}. Another variable CC approach (within our own bubble) 
generically dubbed `holographic dark
energy' (HDE) \cite{4, 5} has also proved to have a potential to shed light  
on both the `old' CC problem and the `cosmic coincidence problem'.

In \cite{4}, the concept of holographic principle, first formulated by 't
Hooft \cite{6} and Susskind \cite{7} as a possible window to quantum
gravity, was attempted to reconcile it with the success of an
effective-quantum-field-theory description of elementary-particle phenomena. 
In order to
prevent formation of black holes within the effective field-theoretical 
description, the entropy for an effective quantum
field theory $\sim $ $L^3 {\Lambda }^3 $, where $L$ is the size of the region
(providing an IR cutoff) and $\Lambda $ is the UV cutoff, should obey the
upper bound \cite{4}
\begin{equation}
L^3 {\Lambda }^3 \leq (S_{BH})^{3/4} \sim L^{3/2} M_{Pl}^{3/2}\;,
\end{equation}
and $S_{BH} \sim L^2 M_{Pl}^2 $ is the holographic 
Bekenstein-Hawking entropy. In an
expanding universe, $\Lambda $ should therefore be promoted to a varying 
quantity, in order for (1) not to  be violated during the course of the
expansion. This gives that a constraint on the maximum energy density in the
effective theory, $\rho_{\Lambda } \sim {\Lambda }^4 $, will be $\rho_{\Lambda
} \leq  L^{-2}M_{Pl}^2 $. Obviously, $\rho_{\Lambda }$ is the energy density
corresponding to a zero-point energy and the cutoff $\Lambda $.    

Let us now show how an alternative derivation of HDE emerges by invoking
the Bekenstein bound \cite{8}. The Bekenstein bound $S_B $ will also 
serve to derive a lower bound on the
HDE-density $\rho_{\Lambda }$. For a macroscopic system in which
self-gravitation effects can be disregarded, the Bekenstein bound is given
by a product of the energy and the linear size of the system, $EL$. In the
context of the effective quantum field theory as described above, it becomes
proportional to ${\Lambda }^4 L^4 $. Note that it is more extensive than the
entropy in the effective field theory. Using the above definitions, one can easily
see that HDE emerges whenever $S_B \leq S_{BH}$, i.e., for  a weakly
gravitating system. Note that this requirement automatically prevents 
formations of black holes, as the Bekenstein bound, in spite of its original
connection with black-hole physics, does not involve the Newton constant. In
addition, from the requirement that the entropy in ordinary quantum field
theory be $\leq S_B $, one obtains (because $S_B $ is more
extensive) 
a lower bound on HDE,
$\rho_{\Lambda } \geq L^{-4}$. Therefore, HDE is constrained both from
above and below \footnote{Note that the lower bound in (2) keeps the
internal consistency of an effective field theory itself  by claiming that
the UV cutoff is  always $\ge $ than the IR cutoff.} 
\begin{equation}
L^{-4} \leq \rho_{\Lambda } \leq L^{-2}M_{Pl}^2 \;.
\end{equation}
The lower bound in (2) should not be considered a surprise but rather a
step towards a resolution of the `new' CC problem, since one knows of no
reason why the CC should vanish.
                 
The main reason why the above HDE model is so appealing in the possible 
description of dark energy is the following:  
When the holographic bound (2) is saturated from  
above, $\rho_{\Lambda }$ gives the right amount of dark energy in the
universe at present, provided $L \simeq H^{-1}$, where $H$ is the Hubble
parameter \footnote{The same is also true  for the future event horizon since
for de Sitter space it coincides with $H^{-1}$.}. 
Moreover, since $\rho_{\Lambda }$ is a running quantity, it also
has a  potential  to shed some light on the `cosmic coincidence problem'. 
On the other hand, 
the most problematic aspect of the saturated HDE model is its compatibility 
with a
transition from decelerated to accelerated expansion. Indeed, as it is well
known, the identification of the IR cutoff with  
the Hubble parameter for spatially flat
universes (as suggested by observations) leads to unsatisfactory 
cosmologies. This is easy to see by
plugging $\rho_{\Lambda } =  L^{-2}M_{Pl}^2 $ (setting a prefactor to unity
for simplicity) into the Friedman equation for flat space 
\begin{equation}
(HL)^2 = \frac{8 \pi }{3} (1 + r)\;,
\end{equation}
where $r=\rho_m /\rho_{\Lambda}$. Thus, a choice $L \sim H^{-1}$ would
require the ratio $r$ to be a constant. This is a general statement, holding
irrespective of whether a fluid is
perfect or not, irrespective of whether $G_N$ is varying or not. 
The interpretation for
various cases is, however, different. For perfect fluids, $r=const.$ means
that the equation of state for dark energy unavoidably  matches that of
pressureless matter, $w =0$ \cite{5}. Thus, we cannot explain the accelerating
expansion of the present universe. For interacting fluids, one is usually
able to
generate accelerated expansion with $r=const.$ as well as to ameliorate the
`cosmic coincidence problem' \cite{9}, but fails to explain that the acceleration era 
set in just recently and was preceded by a deceleration era at 
$z \gsim \, 1$. As a way out, a suggestion to set $L$ by the future event horizon has been
widely accepted \cite{10}.  

In the present paper we argue that a transition from the
matter-dominated era to a dark-energy dominance is left problematic in any saturated 
HDE model, irrespective of the choice for $L$. Then we show on the examples
of three distinct models, two dealing with interacting and one with
noninteracting dark energy, that a transition between the two eras is always
obtained in a nonsaturated HDE model, even for $L \sim H^{-1}$.

The HDE bounds (1) and (2) exhibit that the maximum energy density in the 
effective theory, ${\Lambda }^4 $, is a varying quantity in an expanding 
universe and therefore  cannot dominate other components over the whole
history of the universe. In fact, this is a welcome feature since we know that
in the matter-dominated epoch one should have $\rho_m > {\Lambda }^4 $. The
problem with the saturated HDE lies in the fact 
that during the matter-dominated epoch
this implies that $\rho_m > L^{-2} M_{Pl}^2 $. 
However, then the bound $S_B \leq
S_{BH}$ is violated (we have $S_B > S_{BH}$ during the whole matter
dominance). We do not find it very convincing to have a transition to a
strongly gravitating system when the universe size is decreased by a
factor less than two. We can substantiate this inconsistency by a 
derivation which
closely parallels that of Cohen et.~al.~\cite{4} for an effective theory.
First, we demand 
that the entropy of matter, $S_m \sim (\rho_m /m) L^3 $, be $\lsim $
than the Bekenstein-Hawking bound, $S_{BH}$. However, when $S_{BH}$ is
saturated we necessarily pick up many states with the Schwarzschild radius $R_G
$ much larger than $L$. In the case under consideration, one has $R_G \sim
(mL)L \gg L$ for any plausible mass of dark-matter particles. To avoid these 
difficulties, we propose a stronger bound, $(mL)^{-1} S_{BH}$, which, 
at saturation, gives $R_G \sim L$. This, however, ultimately leads to 
$\rho_m \lsim
L^{-2} M_{Pl}^2$, being thus inconsistent with the matter dominance for the
saturated case.  
  
In the following we parametrize the nonsaturated HDE as \footnote{We 
adopt this parametrization from Ref. \cite{9}, the only paper in which   
Eq. (4) was considered in producing a transition from decelerated  to
accelerated cosmic expansion.}
\begin{equation}
\rho_{\Lambda } = \frac{3}{8 \pi } c^2 (t) L^{-2}M_{Pl}^2 \;,
\end{equation}
and always consider the choice $L=H^{-1}$. With the aid of the Friedman
equation for flat space one can express the function $c(t)$ for such a
choice for $L$ as
\begin{equation}
c^2 (t) = \frac{1}{1 + r} \;.
\end{equation}
The function
$c^2 (t)$ should satisfy $c^2 (a \rightarrow \infty ) \rightarrow 1$
(dark-energy dominance) and $c^2 (t) \ll 1$ during the matter-dominated era.
Note that with the nonsaturated HDE (4), the Friedman equation implies that it
is always the sum $\rho_m + \rho_{\Lambda }$ that  saturates the 
holographic bound  (with $c^2
=1$), i.e.,
\begin{equation}
\rho_m + \rho_{\Lambda } = \frac{3}{8 \pi } L^{-2}M_{Pl}^2 \;.
\end{equation}

In order to encompass a maximal diversity of nonsaturated HDE models, we
consider all three special cases of the generalized equation of
continuity
\begin{equation}
\dot{G}_N (\rho_{\Lambda} + \rho_m) + G_N\dot{\rho}_{\Lambda} +
G_N(\dot{\rho}_m+3H\rho_m) =0 ,
\end{equation}
where overdots denote time derivatives. The special cases stemming from
(7) include: (i) the constant $G_N $, the running $\rho_{\Lambda }$, and the
noncanonical $\rho_m $, (ii) the running $G_N $, the running $\rho_{\Lambda
}$, and the canonical $\rho_m $, and (iii) the constant $\rho_{\Lambda }$,
the running $G_N $, and the
noncanonical $\rho_m $. In this way a distinct
cosmological model is represented by each case, even when the same law for
$\rho_{\Lambda }$ is used in (7) (see below). In the third case, a
specific form for a deviation of $\rho_m $ from its canonical shape
is required as an input dynamics.

As a first example we consider a dynamical  model based on a 
continuous transfer of energy between
the CC and $\rho_m $, which can be adequately described by the 
equation of continuity ($\dot{G}_N =0$, $\rho_m \neq \rho_{m0} a^{-3}$):
\begin{equation}
\dot{\rho }_{\Lambda } + \dot{\rho }_m + 3H\rho_m = 0 \;,
\end{equation}
where the overdots denote time derivatives. In order to solve the 
equations above, an additional assumption regarding the dynamics of 
$\rho_{\Lambda }$ (or $\rho_m $ ) is required. We find it adequate to employ a
decaying CC model \cite{11}, based on the renormalization-group
(RG) evolution for $\rho_{\Lambda }$ and on the choice for the RG scale
$\mu = H$. Here the CC-variation  law is a derivative one, 
thus having a natural appearance of a
nonzero constant in $\rho_{\Lambda }$:
\begin{equation}
\rho_{\Lambda } = C_0 + C_2 {\mu }^2 ,
\end{equation}
where $C_0 $ is  a constant and  $C_2 \lsim M_{Pl }^2 $. With the
{\it ansatz} (9), Eq.~(8), and the Friedman equation,
the following equation for $H(a)$ is obtained: 
\begin{equation}
\frac{H^2}{H_0^2} +\frac{2}{3}\frac{H'Ha}{H_0^2}=
\frac{1+\alpha\left( \frac{H^2}{H_0^2}-1 \right) }{1+r_0} ,
\end{equation}
where $H' \equiv dH/da $ and $\alpha \equiv C_2 H_{0}^2/\rho_{\Lambda 0}$. 
The solutions can be obtained in a closed form as
\begin{eqnarray}
& H(a)=H_0 \sqrt{ 
\displaystyle\frac{1-\alpha+r_0 a^{-3+\epsilon}}{1-\alpha+r_0} } , &
\nonumber \\
& \rho_m=\rho_{m0}a^{-3+\epsilon} , &
\nonumber \\
& \rho_{\Lambda}=\rho_{\Lambda 0} \left[ 1-\displaystyle\frac{\alpha r_0
(1-a^{-3+\epsilon})}{1-\alpha+r_0} \right] , &
\nonumber \\
& r=r_0\displaystyle\frac{a^{-3+\epsilon}}
{ 1-\displaystyle\frac{\alpha r_0(1-a^{-3+\epsilon})}{1-\alpha+r_0} } \; , &
\end{eqnarray}
where 
$\epsilon = 3 \alpha /(1 +r_0 )$. We see that $r(a \rightarrow \infty ) = 0$ 
and therefore $c^2 (a \rightarrow \infty ) =1$. This means that the
holographic bound is saturated asymptotically. From a recent analysis of
density perturbation \cite{12} for the model \cite{11}, one obtains that $C_2
\ll M_{Pl}^2 $ (implying $\alpha \ll 1 $), and therefore we have $c^2 \ll 1 $
in the matter-dominated era.      

Another example involves a transfer of energy between $\rho_{\Lambda }$ and
the gravitational field. The equation of continuity now reads\footnote{For
more on saturated HDE models with variable $G_N$, see \cite{13}.}
\begin{equation}
G_N \dot{\rho}_{\Lambda } + \dot{G}_N (\rho_m + \rho_{\Lambda }) = 0 \;.
\end{equation}
With the same {\it ansatz} (9) for $\rho_{\Lambda }$, we obtain the
following equation for $H(a)$:
\begin{equation}
2\frac{\dot{H}}{H} [ \rho_{m0}a^{-3}+\rho_{\Lambda 0} +c_2(H^2-H_0^2)]
+3\rho_{m0}a^{-3}H =0 ,
\end{equation}
which can be solved numerically. The solution can be used to calculate 
\begin{eqnarray}
& r = \displaystyle\frac{r_0 a^{-3}}
{1+ \alpha \left( \frac{H^2 }{H_{0}^2 } -1 \right)} , &
\nonumber \\
& \displaystyle\frac{G_N}{G_{N0}} = 
H^2 a^3 \displaystyle\frac{r(1+r_0 )}{r_0 (1+r)} . &
\end{eqnarray}
Two limiting cases can, however, be easily read off from Eq. (13). For $a
\rightarrow \infty $, Eq. (13) becomes $\dot{H} = 0$, with the solution $H
\equiv H_{\infty } = {\it const.}$. Therefore again the holographic bound is
saturated asymptotically, $c^2 (a \rightarrow \infty ) =1$.  
On the other hand, for $a \ll 1$, we find that $H
\sim a^{- \frac{3}{2}}$, thus having again $c^2 \ll 1 $
in the matter-dominated era, if $\alpha \ll 1 $. 

The last example is at the same time the most interesting one. It involves a
variable $G_N $ such as to make $\rho_{\Lambda }$ a true constant in (4),
$G_N  \sim H^2 c^2 $. The variation of $G_N $ goes to the expense of a
deviation of $\rho_m $ from its canonical shape, and is described by the
equation of continuity of the type
\begin{equation}
\dot{G}_N(\rho_{\Lambda}+\rho_m)+G_N(\dot{\rho}_m+3H\rho_m)=0 .
\end{equation}
The above model in the saturated version of HDE has been studied
recently in \cite{14}. If we consider only a small deviation from the
canonical $\rho_m $
\begin{equation}
\rho_m=\rho_{m0}a^{-3+\delta }\;,
\end{equation}
where $\delta $ is a constant, one immediately obtains $r = r_0 a^{-3 +
\delta }$ and therefore $c^2 (a \rightarrow \infty ) =1$. On the other
hand, for $a \ll 1$, $c^2 $ goes rapidly to zero. We find an 
explicit expression for
$G_N (a)$
\begin{equation}
G_N(a)=G_{N0} \left(
\displaystyle\frac{r_0^{-1}+a^{-3+\delta }}{r_0^{-1}+1}
\right)^{\delta /(3-\delta )} ,
\end{equation}
in agreement with the present observational bound on $\dot{G}_N /G_N $ if
$\delta \lsim 0.1$. The model can thus be considered as a
``minimal'' dynamical dark-energy scenario as it represents a slight deviation
from the standard $\Lambda $CDM model. What is even more intriguing is the
fact that even the standard $\Lambda $CDM model $(\rho_{\Lambda } = {\it const.
}, \rho_m \sim a^{-3}, \dot{G}_N  = 0)$ can be described as a HDE model
through Eq. (4). If $\dot{G}_N = 0$, a constant $\rho_{\Lambda }$ can
always be parametrized by Eq. (4) as then $H^2 \sim 1 + r$. Hence, even
the standard noninteractive $\Lambda $CDM model does obey all the conditions 
necessary for gaining a status of a (nonsaturated) HDE model. Note that this
is generally true only for the choice $L=H^{-1}$.  

Our final remark concerns the lower bound in Eq. (2). In the case of the
standard $\Lambda $CDM model, the lower bound is violated by $\rho_{\Lambda
}$ at
early times when the temperature is below the Planck temperature but well
above the temperature when nucleosynthesis occurred. For other cases
considered in this paper,
the lower bound is violated at early times when the temperature is well
above the Planck temperature, where the above formulas cannot be expected to
hold.

In conclusion, we have considered several HDE-models in which the saturation 
of the
holographic bound begins  
only with the onset of the dark-energy dominated era. This
provides a consistent description of the transition to the matter-dominated
epoch, when all holographic bound are shown to be respected by the 
matter component. We have shown such behavior not only for the standard HDE
models, but also for a class of `generalized' HDE models in which 
the gravitational coupling is also promoted  
to a time-dependent quantity. Thus it is a generic
feature of nonsaturated HDE models to provide a transition from a
decelerated to an accelerated era, even for flat space and the choice of the
IR cutoff in the form of the Hubble parameter. We have shown that,
generically, even the noninteracting standard $\Lambda $CDM model fits
within this nonsaturated HDE description. It is also shown that an alternative
derivation of HDE is possible by invoking the Bekenstein bound. In this
case,
HDE is also supplied with the lower bound, which is shown to be violated
only at very early times in the universe evolution, thus signaling a
breakdown of the effective-field-theory description and/or a transition to a
system with strong self-gravity. Our study therefore retains HDE as a
plausible candidate for the dark energy of the universe.        

{\bf Acknowledgments.} This work was supported by the Ministry of Science,
Education and Sport
of the Republic of Croatia, and
partially supported through the Agreement between the Astrophysical
Sector, S.I.S.S.A., and the Particle Physics and Cosmology Group, RBI.

\end{document}